# Generative Artificial Intelligence in Scientific Research: Individual Benefits, Collective Risks, and a Framework for Responsible Research with AI

Fulvio Castellacci[1], Tommaso Ciarli[2], Yuan Gao[3], Marianna Marino[4], Giacomo Marzi[4*], Massimo Riccaboni[4], Maria Savona[5], Simone Vannuccini[6]

*Author order is alphabetical. *Corresponding author: Giacomo Marzi, Scuola IMT Alti Studi Lucca,* *giacomo.marzi@imtlucca.it*

## Abstract

This paper examines the tension between the benefits of generative artificial intelligence (AI) for scientific research and the unresolved governance questions that accompany its rapid adoption. Drawing on an academic roundtable held at the AI for Science and Innovation Workshop (Scuola IMT Alti Studi Lucca, April 2026) and on a fast-expanding empirical literature, it maps the disagreement within the research community across four stages of the research process: funding, research tasks, publication and peer review, and use and uptake. The empirical case for AI's productivity, augmentation, and democratization effects has strengthened. The picture changes once productivity is disaggregated: AI-assisted work shows measurable gains in publication volume and citation share, while the evidence on novelty, disruption, and breakthrough output remains ambiguous or negative. We argue that the divergence between private and social returns arises through three analytically distinct mechanisms, namely information asymmetry, negative externalities on a shared knowledge base, and depletion of research capacity, and that each calls for a different governance instrument. We propose Responsible Research with AI (RRAI), an extension of the Responsible Research and Innovation tradition organized around four principles that operate at different levels of the research system: disclosure, differentiation, narrative, and proportionality. RRAI builds on existing institutional scaffolding, including the EU AI Act, UNESCO, and the OECD, and aims to preserve AI's productivity gains while addressing systemic risks that individual researchers can neither observe nor manage on their own.

---

[1] TIK Centre for Technology, Innovation and Culture, University of Oslo, Oslo, Norway
[2] UNU-MERIT, United Nations University, Netherlands; SPRU, University of Sussex, United Kingdom
[3] University of East Anglia, Norwich Research Park, United Kingdom
[4] IMT School for Advanced Studies Lucca, Lucca, Italy
[5] University of Sussex, Brighton, United Kingdom
[6] Université Côte d'Azur, CNRS, GREDEG, France

## 1. Introduction

Generative artificial intelligence (AI) has entered scientific research practice at a pace that few anticipated. Researchers now use large language models to draft manuscripts, generate and debug code, synthesize literature, design experiments, formulate hypotheses, and stress-test manuscripts before submission. Bibliometric evidence shows that AI methods had diffused across scientific domains well before the public release of generative models in November 2022 (Bianchini et al., 2022). The prospect of an "AI scientist" running a closed-loop research pipeline no longer reads as remote speculation (Lu et al., 2026).

The ethical, procedural, organizational, and governance implications of this adoption are wide and unresolved. Leading journals and funding agencies may not be able to absorb the increase in submissions (Rees and Wilsdon, 2026), and they are issuing heterogeneous policies on AI use. The first AI-related retractions are emerging, and there is no consensus on whether existing self-regulatory mechanisms are adequate (Resnik and Hosseini, 2025). Ethical breaches in peer review, the activity on which the integrity of the entire system rests, may be the more immediate concern (Gartenberg et al., 2026).

This paper addresses a selected set of these issues. It originates from an academic roundtable held at the AI for Science and Innovation Workshop (Scuola IMT Alti Studi Lucca, 8–9 April 2026), where researchers, primarily economists and innovation scholars, debated the motion that "the benefits of generative AI adoption in scientific research outweigh its risks, and existing self-regulatory mechanisms within academia are sufficient to manage them." The exercise was structured as a point-counterpoint, with each side presenting the strongest available evidence. A pre-debate vote of ten in favor and six against became nine in favor and eight against afterwards. The shift is small and the group was not representative, but its direction is informative. Participants who changed position accepted the claim about benefits and rejected the claim about the sufficiency of self-regulation. The compound structure of the motion is where the disagreement in the field actually lies, and it is the structure this paper follows.

We consider four stages of the research process: (i) research funding, including proposal writing, evaluation, and allocation; (ii) research tasks, including literature synthesis, hypothesis formulation, coding, data analysis, and drafting; (iii) publication and peer review, including manuscript preparation, editorial triage, and reviewer assessment; and (iv) research use and uptake, including dissemination, replication, application, and learning.

Recent contributions have examined AI in scientific research from several angles. Resnik and Hosseini (2025) develop an ethics framework centered on disclosure and authorship. Messeri and Crockett (2024) identify mechanisms through which AI generates illusions of understanding. Van Quaquebeke et al. (2025) examine how AI reshapes scientific inquiry beyond efficiency gains. Bender et al. (2021), Bommasani et al. (2021), and Birhane et al. (2024) analyze the systemic biases and societal effects of foundation models. Gartenberg et al. (2026) address the crisis of incentives in peer review.

Our contribution is twofold. First, we specify the conflict between private and social returns to AI use in research. The literature frequently invokes a divergence between individual and collective outcomes without distinguishing the mechanisms that produce it, and the distinction matters because different mechanisms require different instruments. We identify three. Information asymmetry arises where AI-assisted output cannot be distinguished from unassisted output, so that evaluators lose the ability to infer future capability from past performance (Section 2.2). Negative externalities on a shared knowledge base arise where individually rational AI use narrows the range of topics studied, homogenizes framing, and degrades the corpus on which subsequent research and subsequent models both depend (Sections 2.1 and 2.7). Capacity depletion arises where the tasks AI absorbs are the same tasks through which research capability is acquired (Section 2.4). Workload and funding effects (Section 2.3), the historical record of self-regulation (Section 2.5), and the architecture of attribution (Section 2.6) condition how these three mechanisms operate.

Second, we propose Responsible Research with AI (RRAI) as a governance framework that extends the Responsible Research and Innovation tradition (Owen et al., 2012; Stilgoe et al., 2013; Von Schomberg, 2013) and that matches instruments to mechanisms. RRAI is organized around four principles that operate at different levels of the research system: disclosure at the informational level, differentiation at the pedagogical level, narrative at the communicative level, and proportionality at the distributional level.

The paper draws on the economics of innovation, the science of science, and science and technology studies. Section 2 assesses the empirical case for AI's benefits and the limits of that case. Section 3 develops the RRAI framework. Section 4 concludes.

## 2. Individual benefits and collective risks

### 2.1. Research productivity, reconsidered

Research productivity has been argued to have been declining for decades: ideas are getting harder to find as the frontier expands and the stock of accumulated knowledge grows (Bloom et al., 2020; Jones, 2009). The claim is disputed, on the grounds that ideas are becoming cheaper to produce, that the measurement is flawed, or that the pattern reflects declining technological disruptiveness, not lower research efficiency, and these alternative explanations have themselves been criticized (Arza et al., 2026). The literature identifies barriers that bear directly on the AI claim: an expanding frontier requires more knowledge to be absorbed before a contribution becomes possible (Jones, 2009), disciplinary fragmentation impedes recombination (Uzzi et al., 2013), and research organization has become more team-based, longer-cycle, and risk-averse (Wuchty et al., 2007; Wu et al., 2019).

AI could in principle address several of these barriers by reducing the marginal cost of navigating the knowledge space and by enabling smaller teams to undertake work that previously required larger collaborations (Bianchini et al., 2022; Cockburn et al., 2018), a mechanism formalized in models of prioritized search (Agrawal et al., 2024) and of an effectively enlarged idea-generating population (Kremer, 1993). The question is therefore not whether AI raises individual output but whether the indicators that correlate with AI adoption capture the social value of research. Kapoor and Narayanan (2025) describe the divergence as a production-progress gap: publication output has grown by orders of magnitude over the past century while disruption, breakthrough, and recognition measures have flattened, and AI, by accelerating production without addressing the underlying bottlenecks, is likely to widen the gap, not close it. What is at stake is whether time-saving gains at the individual level are sustainable in terms of novelty, quality, and the accumulation of propositional rather than merely prescriptive knowledge, in Mokyr's (2002) sense.

The literature offers at least five ways of measuring research productivity, and they do not move together. Publication output per researcher has risen sharply: Kusumegi et al. (2025), analyzing 2.1 million preprints, report increases ranging from 23.7 to 89.3 percent depending on scientific field and author background, with the largest effects in the social sciences, the humanities, and among non-native English speakers. Citation share has shifted toward AI-assisted work (Chugunova et al., 2026; Trisović et al., 2025), and experimental productivity in writing and coding tasks shows clear gains concentrated among less-skilled workers (Noy and Zhang, 2023; Brynjolfsson et al., 2025). Novelty and disruption show no analogous gains (Park et al., 2023), and breakthrough output has not accelerated commensurately with publication volume (Pammolli et al., 2011; Bhattacharya and Packalen, 2020). Table 1 reports the relevant magnitudes.

Kusumegi et al. (2025) also report that LLM adoption has inverted the historical relationship between linguistic complexity and assessed quality, producing an influx of manuscripts that are more complex and substantively weaker. The governance problem sits in that gap. Academic evaluation counts most easily the indicators that rise under AI adoption, and the indicators that do not rise are the ones carrying social value. AI may also change what can be asked, since large language models can act as surrogates in social science experiments, although their fidelity as substitutes for human subjects remains contested (Gao et al., 2025; Grossmann et al., 2023), and most surveyed researchers regard AI as transformative (Van Noorden and Perkel, 2023; Chugunova et al., 2026).

The collective picture is less favorable. Individual creativity gains coincide with a reduction in the collective diversity of novel content (Doshi and Hauser, 2024), and AI may create what Messeri and Crockett (2024) call illusions of understanding, in which researchers believe they comprehend a domain more thoroughly than they do because AI removes the friction that would otherwise signal gaps in knowledge. The most consequential risk is a homogenization feedback loop. Models trained on existing corpora encode their patterns and conventions; researchers receive outputs reproducing those patterns; AI-assisted papers then enter the literature and become training data for the next generation of models. The process is invisible to any individual researcher, because each instance of assistance feels neutral and useful. Shumailov et al. (2024) demonstrate the technical mechanism in what they term model collapse, and Song et al. (2025) provide upstream evidence, finding pairwise error correlations above 0.8 in chemistry and physics across frontier models from four providers and attributing the convergence to shared pre-training corpora, not to architecture.

Hao et al. (2026), analyzing 41.3 million papers across the natural sciences, document both sides of the ledger in one design. Researchers engaged in AI-augmented work publish 3.02 times more papers and receive 4.84 times more citations, while AI adoption shrinks the collective volume of topics studied by 4.63 percent and reduces engagement between scientists by 22 percent, as AI-augmented work moves toward the areas richest in data.

Individual gains are well evidenced. Acceleration at the level of the system is not. If AI adoption concentrates scientific attention on computationally tractable questions, researchers working on data-sparse but socially important problems face a structural disadvantage.

## 2.2. Selection and adverse selection

If AI use is not disclosed, and if AI-assisted output cannot readily be distinguished from unassisted output, then hiring committees, tenure boards, journal editors, and grant panels face an information asymmetry structurally equivalent to the one Akerlof (1970) described. Selection may favor skilled users of generative AI over skilled researchers, and the distinction cannot easily be drawn.

The pro-AI response is that if the output is better, the method of its production is secondary. In competitive chess, computer assistance counts as doping because the game exists to evaluate unaided strategic reasoning, whereas science exists to produce knowledge. The objection holds only partially. Where AI output emerges from a process that cannot be reconstructed, there is no way to verify that an insight was generated through a traceable and reproducible route.
Evaluating the novelty and quality of an output is also distinct from allocating credit to those who produced it, and academic institutions do both. Hiring, tenure, promotion, and grant decisions are simultaneously assessments of past output and predictions of future capability. If evaluators cannot separate the human from the machine contribution, past output loses predictive value for future capability, which is the signal these decisions actually require. The same applies to recommendation letters, whose informational content falls if they are themselves AI-generated.

The empirical conditions for this dynamic are present. Only 1.5 percent of medical manuscripts disclose LLM involvement, and revealed AI assistance reduces perceived manuscript quality by 0.32 points on a five-point scale, which is a direct incentive against disclosure (BaHammam, 2025). Estimates of undisclosed prevalence vary widely across detection methods and corpora (Kobak et al., 2025; Liang et al., 2024), and that dispersion is itself a governance problem: the research system has no reliable measure of how much of its own output is AI-mediated, yet it is being asked to make selection decisions in that condition. Detection is also harder than it appears. Lu et al. (2026) report a fully AI-generated manuscript achieving peer-review scores sufficient for acceptance at a 2025 machine learning workshop, and an audit of 2.5 million PubMed-indexed papers found one fabricated reference per 277 papers in the first weeks of 2026, up from one per 2,828 in 2023.[2]

Voluntary norms cannot correct an asymmetry of this kind, because the private cost of disclosure is positive while its benefits accrue to the system as a whole. What academic selection requires, on this reading, is disclosure that is mandatory, standardized, and verifiable.

### 2.3. Workload and the political economy of research funding

Two questions are connected here: how AI reshapes research tasks, and how that reshaping is interpreted by those who fund research. We treat them together because the fiscal consequences follow directly from how the labor effects are read.

Grafström (2025) decomposes academic workload into automatable tasks, where AI substitutes for labor; non-automatable tasks, which require human judgment; and AI-induced tasks such as verification, compliance, and output curation. AI reduces effort on the first, expands the second, and creates the third, so the net effect is ambiguous and, in the model, likely positive: researchers work more, not less (see also Agrawal et al., 2026). Competitive escalation follows, because as AI assistance becomes standard the expectation of AI-augmented output rises with it. Evidence from workplace settings outside academia points to a related loss, since recipients of low-quality AI-generated content report spending substantial time correcting it, so that gains at the point of production are offset downstream (Niederhoffer et al., 2025). The assumptions may not hold uniformly, and where publication pressure is lower AI may reduce workload. At the macroeconomic level, the estimated contribution of AI to total factor productivity is 0.66 percent over a decade (Acemoglu, 2025), consistent with task-based models in which the net effect depends on the balance between displacement and reinstatement (Acemoglu and Restrepo, 2019).

[2] See https://www.nursing.columbia.edu/news/nearly-3-000-peer-reviewed-medical-papers-have-fake-citations-columbia-nursing-ai-assisted-audit-finds

The labor-level finding has a direct fiscal implication. If academics demonstrate that AI can absorb a substantial share of research tasks, policymakers may reduce research budgets. Public research spending is not politically protected in the way that health or defense spending is, precisely because attributing outcomes to research is harder than in other domains of public expenditure.[3] AI worsens the attribution problem on two fronts. By increasing publication volume without commensurate gains in demonstrable outcomes, it makes the gap between indicators and outcomes more visible. And by entering proposal writing, it degrades the proposal as a signal of research capability, leaving funders either to develop new evaluation strategies or to accept more noise in allocation. The research community therefore has a shared interest in a more accurate public account: AI redistributes research tasks and expands verification work, and does not reduce total labor input.

### 2.4. Doctoral training and capacity depletion

Doctoral training operates through apprenticeship. Early-career researchers improve writing by writing, learn to code by debugging, and struggle with statistical output until the patterns become interpretable. AI can accelerate the productive component of these tasks while short-circuiting their learning function (Van Quaquebeke et al., 2025), and may therefore democratize access to the artifacts of research without democratizing the capability to produce it.

---

[3] On the vulnerability of public research budgets to political reallocation, see https://www.brennancenter.org/our-work/research-reports/cost-trump-administrations-attacks-research-funding

Against this, the same concern has attached to every previous tool and has never been borne out: calculators did not eliminate mathematical reasoning, spell-checkers did not eliminate literacy, and statistical software did not eliminate quantitative judgment. The analogy may not hold. A calculator automates a narrow procedural task without touching the conceptual architecture that makes the task meaningful, whereas generative AI operates on the cognitive processes themselves. Writing, literature synthesis, hypothesis generation, and data analysis are not peripheral activities for a doctoral student; they are the substance of the training years (Van Quaquebeke et al., 2025).

The available evidence points to mode of engagement rather than use as the relevant variable. Lehmann et al. (2025), in two pre-registered experiments in which subjects learned to program with or without access to a large language model, find no difference in average learning outcomes but a substantial difference by usage pattern: students who used the model to generate solutions covered more material while understanding less of it, whereas those who used it to request explanations deepened their understanding. Kosmyna et al. (2025) report converging evidence from a smaller neuroimaging study of essay writing, in which participants using an LLM showed weaker brain connectivity, lower recall of their own text, and a reduced sense of ownership, with effects persisting after the tool was withdrawn. That study has a small sample and remains a preprint whose interpretation has been questioned, so it should be read as suggestive rather than settled.

Bjork and Bjork's (2011) framework of desirable difficulties explains the pattern. The friction of writing a poor first draft or debugging without assistance is the mechanism through which the tacit knowledge of a discipline is internalized, so removing it improves the immediate artifact while preventing the consolidation that makes independent work possible later. Students who interrogate AI output reinstate the difficulty; students who copy eliminate it. Without institutional guidance, however, the rational strategy for a doctoral student under resource constraints is to copy, and simulation of the publication pipeline under an AI-driven submission surge shows the downstream consequence: early adopters gain, but as volumes rise acceptance rates fall for everyone, with junior faculty at less prestigious institutions disproportionately disadvantaged (Jiang, 2025; see also Beugelsdijk and Bird, 2025).

This mechanism is neither an externality nor an asymmetry but a depletion of capacity, and the instrument it calls for is correspondingly different. Neither disclosure nor prohibition addresses it. Differentiated pedagogical guidance organized around mode of engagement does.

### 2.5. Self-regulation: swift response, stalled implementation

Since 2023, every major publisher, including Springer Nature, Elsevier, and Wiley, together with the Committee on Publication Ethics, has issued AI disclosure requirements, within months of the release of ChatGPT and faster than any previous governance adaptation in academic publishing. Speed of response has not translated into effective governance. Three years on, disclosure rates in the fields where they have been measured remain between one and two percent despite widespread use, because the incentive structure penalizes disclosure (BaHammam, 2025).

The record of analogous self-regulation should be read with care. Data-sharing requirements remain irregularly complied with despite the FAIR principles (Wilkinson et al., 2016), and pre-registration remains a minority practice outside experimental psychology (Ioannidis, 2005; Nosek et al., 2015). Voluntary norms coordinate effectively when the costs of defection are visible and the benefits of compliance are immediate and individual (Ostrom, 1990). In the case of AI use, the costs of non-disclosure are diffuse and collectively borne, in the form of degraded selection signals, weakened peer review, and reduced public trust, while the benefits of disclosure carry individual reputational risk. Non-disclosure therefore has the structure of a commons problem, not of an ethical failure, which is why exhortation has not worked. AI governance more generally shows the same pattern at larger scale (Jobin et al., 2019).

Proponents of self-regulation respond that AI is different in kind, that the stakes are higher and the community more motivated, and that the speed of the initial publisher response demonstrates a capacity for rapid adaptation. The response has merit, but researchers themselves are not confident in it. In a survey of 6,215 researchers, 85 percent want regulatory guidance, 58.7 percent want it specifically from supranational institutions, and 70 percent identify legal uncertainty as the principal barrier to adoption (Chugunova et al., 2026). The demand for governance is coming from inside the research community. At the same time, professional norms remain flexible and field-sensitive in a way that centralized rules are not, and the administrative burden of centralized rules falls unevenly across institutions. Neither instrument works alone. Institutional governance should augment self-regulation and should target the specific incentive structure that makes disclosure individually costly where it is collectively valuable.

### 2.6. Attribution and the incentive structure of research

A distinct question is how AI affects the systems through which intellectual contributions are owned, attributed, and rewarded. These matter because the incentive structure of academic research depends on a relatively stable mapping between authorship, originality, and credit, and generative AI disrupts that mapping in three ways.

First, authorship attribution has become contested. Editorial bodies have ruled that AI systems cannot be listed as authors because they cannot be held accountable for errors (COPE, 2023), which is defensible but leaves unresolved how to attribute the human contribution when substantial portions of a manuscript are machine-generated (Stokel-Walker, 2023). A further scenario is one in which the research paper ceases to be a fixed artifact and becomes a continuously updated, AI-enabled object whose derivative versions diverge progressively from any original, a trajectory already visible in computational notebooks and versioned preprint platforms (Fraser et al., 2021; Perkel, 2018; Rule et al., 2019; Stodden et al., 2016). If it materializes even partially, the citation graph, the patent system, and the criteria for authorship all lose the stable identifiable object on which they depend.

Second, patent law has had to confront AI inventorship. The DABUS cases, in which an AI system was named as inventor on applications filed in multiple jurisdictions, produced consistent rulings that only natural persons can be inventors under existing law (Abbott, 2020; Thaler v. Vidal, 2022). The outcome is legally settled but leaves a practical tension: if AI systems generate patentable inventions that cannot be patented as such, the result is either unprotected innovation, which weakens incentives to invest, or strategic non-disclosure, which weakens diffusion.

Third, idea attribution and the citation graph are under structural pressure. Generative models synthesize patterns across large numbers of training documents without preserving citation links to sources, so work produced with AI assistance may incorporate ideas whose originators go uncredited. Researchers have always built on prior work without exhaustive citation, so the problem is not new in degree. It is new in kind, because AI mediation breaks the chain of attribution in a way that human writing did not, and the reward system of science has rested on that chain remaining traceable since the rise of citation indexing (Garfield, 1979). Tools for detecting and sourcing AI-mediated text are emerging (Kobak et al., 2025; Liang et al., 2024), but the tension between generative capacity and reliance on traceable provenance is unresolved.

Less visible is what happens to incentives once attribution becomes systematically less reliable. Endogenous growth models assume that researchers capture some private return on intellectual investment, with the residual spillover constituting social value (Romer, 1990). If AI mediation erodes the ability to identify and reward original contributions, the reputational return falls and the standard model predicts under-investment in original work, with the optimization target shifting from producing new ideas toward recombining existing ones. Attribution is therefore not addressable by disclosure alone: it requires AI-aware citation standards and new categories of contribution distinguishing original framing from AI-assisted execution.

**2.7. Distributional effects**

The democratization argument deserves weight. AI can lower entry barriers for non-native English speakers, reduce dependence on costly proofreading services that have functioned as an informal publication barrier in lower-income countries, and allow scholars at resource-constrained institutions to compete on the strength of their ideas (Chugunova et al., 2026), and Kusumegi et al. (2025) find the largest output gains precisely among non-native English speakers. But lowering the cost of producing a polished draft does not by itself raise the capability to do research, and the evidence on whether the gains are being realized is qualified along four dimensions.

Geography. An analysis of more than 230,000 articles finds that authors from the Global South adopt AI writing tools more aggressively while Western authors continue to accumulate more citations, with only modest increases in the visibility of non-Western work (Khan et al., 2025). Because AI adoption elevates research in data-rich, computationally legible areas (Hao et al., 2026), researchers whose questions concern data-sparse problems face a compounding disadvantage, and this group disproportionately includes scholars in lower-income countries working on context-specific questions, qualitative social scientists, and humanities researchers.

Gender. A persistent gender gap in AI use has been documented, with approximately 71 percent of the gap explained by differential familiarity with the technology rather than by attitudinal resistance (Chugunova et al., 2026; Carvajal et al., 2024). Where AI use compounds existing differences in publication rates, the gap in academic visibility may widen.

Discipline and topic. Trisović et al. (2025) report 2024 adoption rates of 34 percent in linguistics, 18 percent in computer science, and 4.6 percent in engineering, with most other fields between 1 and 3 percent. Fields that adopt more aggressively may publish more and therefore attract more funding, and the feedback loop between AI compatibility and resource allocation concentrates effort in areas that are already favored. Whether that is socially beneficial depends on whether AI-favored topics match the topics society needs research on, which is an open and contested question.

Infrastructure. The most consequential distributional concern is structural, and it is the one the equalizer narrative obscures. The infrastructure on which generative AI depends, namely large-scale training data, computational resources, model weights, and publishing analytics, is concentrated in a small number of high-income countries and a smaller number of corporate actors within them, and science and technology studies have long argued that research infrastructures shape what is knowable and for whom (Edwards, 2010; Jasanoff, 2004). Measurement supports the reading: Trisović et al. (2025) find that the median model adopted by scientists in 2024 was 26 times smaller than the median model being built that year, up from 1.8 times in 2019, and that adoption of closed models is growing at 336 percent annually against 20 percent for open-weight models. Since researchers using larger models publish in higher-impact venues, access asymmetry has begun to translate into outcome asymmetry, and foundation models also encode the priorities and blind spots of their developers (Bender et al., 2021; Birhane et al., 2024). The democratization argument assumes that AI access is the binding constraint; the structural argument holds that access is itself shaped by asymmetries that predate generative AI and that AI may amplify rather than redress.

The distributional evidence therefore qualifies the equalizer claim without refuting it. AI removes one barrier, in language and drafting cost, while reinforcing others, in data availability, infrastructure, and topic selection. Governance that ignores this asymmetry will produce compliance burdens that fall hardest where capacity is weakest.

### 2.8. The normative fault line: science or scientists

A tension runs through the whole of the preceding analysis. It concerns whether scientific institutions exist primarily to produce knowledge or also to develop human capacity, certify expertise, and maintain a system of merit-based allocation, and the governance architecture appropriate to the research system depends on the answer.

An output-maximizing view supports minimal regulation. On this reading the selection problem in Section 2.2 is secondary, the capacity problem in Section 2.4 matters only insofar as learning affects output, and governance should concern output quality alone. A capacity-development view supports differentiated regulation. If institutions also train researchers and certify expertise, then substituting AI for human judgment introduces costs that output metrics do not register: illusions of understanding at the individual level (Messeri and Crockett, 2024), the displacement of human accountability in science communication, on which public trust in expertise partly rests (Silva Luna et al., 2025), and the narrowing of the knowledge frontier that becomes visible only retrospectively.

The choice cannot be settled on empirical grounds, and the pace of adoption makes deliberation urgent. The pragmatic implication is that governance design should be explicit rather than implicit about which view it adopts, and should preserve the capacity-development functions of academic institutions even where output-maximizing logic would favor unrestricted adoption. The framework in Section 3 adopts the second view.

## 3. Toward a governance architecture: the RRAI framework

### 3.1. Design logic and institutional anchors

Section 2 identifies three mechanisms through which private and social returns to AI use in research diverge, and shows that they are structurally different. Information asymmetry is a problem of legibility, and it calls for instruments that make the informational environment observable. Negative externalities on a shared knowledge base constitute a commons problem, and they call for instruments that monitor and internalize collective effects. Capacity depletion is a dynamic human-capital problem, and it calls for instruments that operate on training, not on output. A single instrument applied uniformly will address at most one of the three, which is why the current debate between unrestricted adoption and top-down regulation is poorly specified.

We propose Responsible Research with AI (RRAI) as a way to organize the design effort. RRAI extends the Responsible Research and Innovation literature (Owen et al., 2012; Stilgoe et al., 2013; Von Schomberg, 2013) to the specific case of generative AI in scientific research. The extension does more than transfer an existing framework to a new object. RRI was developed for the governance of research directed at society, and its dimensions of anticipation, reflexivity, inclusion, and responsiveness address primarily the content and direction of research agendas. What generative AI disrupts is different in kind: it is the internal machinery through which the research system evaluates its own members, allocates credit, and reproduces capability. RRAI retains the RRI commitment to reflexive and inclusive governance and specifies instruments for that internal machinery.

The correspondence to the RRI dimensions is direct. Anticipation is operationalized as monitoring, through the compliance and effect indicators set out in Table 2. Reflexivity is operationalized as disclosure, which requires researchers to make their own process visible to themselves as well as to evaluators. Inclusion is operationalized as proportionality, which imposes the condition that governance must not exclude the least-resourced participants. Responsiveness is operationalized as differentiation, which requires that requirements adjust to career stage and to mode of engagement instead of applying uniformly.

RRAI builds on existing institutional scaffolding instead of proposing rules without precedent, and the anchors are selected on three criteria. They must be binding or capable of becoming binding, they must operate above the level of the individual publisher, and they must already contain a mechanism that is transferable to research practice. The EU AI Act (European Parliament and Council, 2024; Regulation (EU) 2024/1689) qualifies because it is binding and because its tiered, risk-proportionate transparency obligations transfer directly. The UNESCO Recommendation on the Ethics of Artificial Intelligence (UNESCO, 2021) qualifies because it is supranational and addresses scientific research explicitly. The OECD AI Principles (OECD, 2019) qualify because they supply the accountability and transparency vocabulary that national funders have already adopted. Emerging guidelines from national research funders, including the German DFG and the French ANR, qualify because funders can attach conditions to grants, which makes their guidance enforceable in a way that publisher policy is not. Cross-country monitoring instruments such as the Global AI Index (2026) supply the capacity data required to calibrate proportionality. We do not treat the general AI ethics literature as an anchor, because that literature consists overwhelmingly of voluntary guidelines with almost no binding force (Jobin et al., 2019), which is precisely the condition RRAI is designed to move beyond.

### 3.2. Four principles at four levels

The four RRAI principles operate at different levels of the research system. The differences in level are deliberate, and they are the reason the four cannot be collapsed into a single compliance requirement. Disclosure operates at the informational level and addresses adverse selection. Differentiation operates at the pedagogical level and addresses capacity depletion. Narrative operates at the communicative level and addresses the fiscal externality that arises from how AI's labor effects are represented to funders and to the public. Proportionality operates at the distributional level and constrains the design of the other three, so that compliance costs do not fall regressively. Figure 1 summarizes the relations among them.

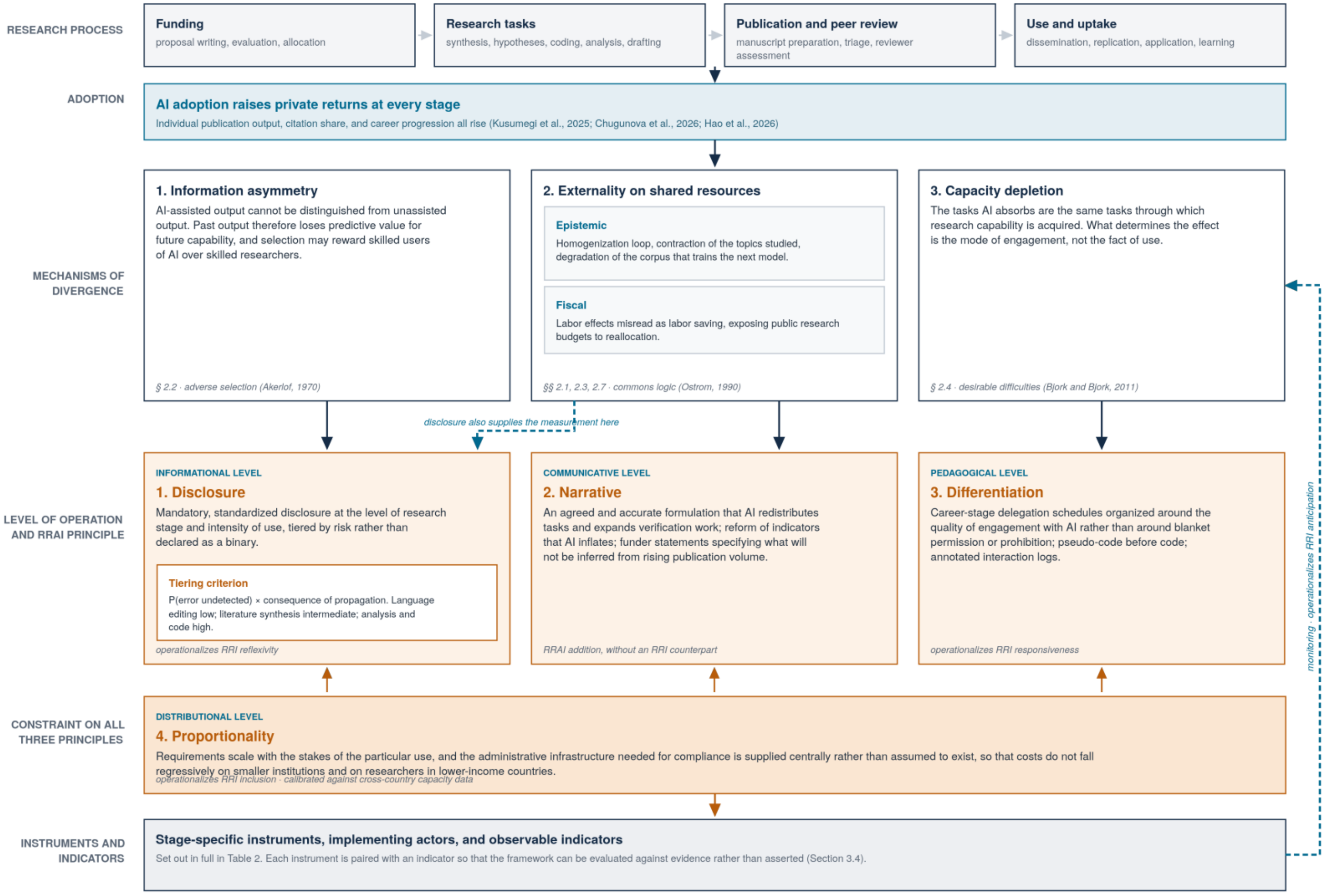

**Figure 1.** The RRAI framework: three mechanisms of divergence between private and social returns, the four principles that address them at four levels of the research system, and the monitoring loop connecting instruments back to mechanisms.

### *Disclosure*

The adverse selection argument developed in Section 2.2 implies that disclosure of AI use is a condition for trust in the academic selection system, and that voluntary disclosure under adverse selection does not converge on a workable standard (Akerlof, 1970). Three design questions have been sidestepped in the current debate.

The first is granularity. A binary declaration of whether AI was used is almost certainly inadequate, because it does not distinguish a researcher who used AI to polish English from one who used AI to generate the statistical analysis. Stage-level disclosure is considerably more informative.

The second is intensity. Completing a literature review with AI assistance differs categorically from having a literature review generated. A workable standard must capture this gradient without becoming so burdensome that compliance collapses.

The third is the tiering criterion, and this is where the analogy with the EU AI Act has to be made precise rather than invoked. The Act tiers obligations by risk, understood as the probability of harm multiplied by its severity, and imposes transparency duties in proportion. The research equivalent is the probability that an AI-induced error passes undetected, multiplied by the consequence of that error propagating into the literature. Language editing scores low on both terms, since errors are visible to the author and their consequences are stylistic. Statistical analysis and code generation score high on both, since errors are not visible without independent replication and their consequences propagate into every downstream citation. Literature synthesis sits between the two, because fabricated references are detectable while false attributions of ideas are not. This criterion, and not a general appeal to tiering, generates the disclosure schedule in Table 2.

Disclosure then operates differently at each stage of the research process. At the funding stage, proposal-writing assistance is sufficiently widespread that funders need explicit policies on acceptable use and on what must be declared within the proposal itself. At the research task stage, disclosure should distinguish among literature review, data analysis, code generation, and drafting. At the publication stage, a methods of AI use statement should accompany the manuscript, analogous to the methods sections that already exist for data and analysis. At the peer review stage, the situation is least tractable. Many journals explicitly forbid reviewer use of AI, which makes disclosure structurally impossible: a reviewer who used AI cannot acknowledge it without violating policy, and a reviewer who did not cannot demonstrate the absence. One concrete alternative is a tiered Reviewer AI Use Statement, structured like the schedule proposed for authors and distinguishing language checking, literature retrieval, drafting of reviewer comments, and assessment of technical validity. This would make reviewer AI use trackable through the editorial process and would generate the data needed to monitor AI's effects on peer review quality over time. Whether prohibition or disclosure-with-tiering produces the more reliable outcome is an open empirical question, and it is one that the editorial data generated under a tiering regime would answer.

The deeper design challenge is normative reframing. A system in which disclosure reduces perceived manuscript quality by 0.32 points (BaHammam, 2025) will not produce truthful disclosure at any level of mandate, because researchers will comply minimally where compliance is punished. The analogy with clinical trial pre-registration is informative. Pre-registration was initially resisted as an admission of untrustworthiness and became a marker of methodological quality over roughly a decade, largely because leading journals began to require it and to display it. The same trajectory is available for AI disclosure, but only if journal policy, evaluation criteria, and professional norms actively support the inversion. Disclosure does not restrict AI use, and a researcher who uses AI transparently and produces strong work should face no penalty. Achieving the inversion from confession to signal of rigor requires coordinated action at the level of the journals that set prestige.

***Differentiation***

Section 2.4 established that the effect of AI on research capability depends on how AI is used and not on whether it is used, and that the effect differs by career stage. Differentiation follows directly. Requirements should vary by career stage and by mode of engagement, not take the form of a uniform permission or a uniform prohibition.

Doctoral programs should specify which tasks may be delegated at each stage of training, with the developmental function of apprenticeship as the explicit design objective. Practical elements are available now and are implementable at departmental level, which makes this the most tractable component of RRAI and a natural site for near-term experimentation. They include requiring students to draft pseudo-code before functional code, so that command of the underlying program logic is demonstrated independently of its implementation; requiring annotated logs of AI interaction to be submitted alongside written work, which converts private use into a reviewable artifact and shifts the incentive from copying toward interrogation; introducing a methods of AI use section in doctoral theses; and building seminars on the critical evaluation of AI output into doctoral curricula, on the grounds that the returns to verification skill rise as the volume of plausible unverified text rises.

Differentiation extends beyond doctoral training. Supervisors need guidance on what to require of students, editors need guidance on what to expect of authors at different career stages, and evaluation committees need to know how to read an AI use statement without penalizing candor. A regime that mandates disclosure without differentiating expectations by career stage would place the heaviest verification burden on those least equipped to carry it.

***Narrative***

Section 2.3 showed that AI changes the distribution of tasks within the research process rather than reducing total labor input (Grafström, 2025), and that the simplified account of AI as labor-saving carries a fiscal risk for public research funding. Narrative belongs in a governance framework rather than in a communications appendix, for two reasons. The representation of AI's labor effects has budgetary consequences, and the research community is the only actor positioned to correct that representation.

Three operational commitments follow. First, professional associations and research organizations should adopt a shared and accurate formulation in submissions to funders and in public communication: AI redistributes research tasks, expands verification work, and does not reduce total labor input. Second, indicator reform should accompany the formulation. Where research assessment counts publication and citation volume, AI adoption inflates precisely the measures that are becoming less informative about outcomes, and funders who rely on them will draw the wrong inference. Assessment frameworks should therefore give explicit weight to indicators that AI does not inflate, including replication, data and code availability, and evidence of use outside academia. Third, funders should be asked to state in advance what they will and will not infer from a rise in publication volume in AI-adopting fields. A public commitment of this kind is a low-cost protection against the inference that measured productivity gains justify budget reductions.

Narrative is the principle most easily dismissed as advocacy. It is included because the alternative is a governance framework that manages research integrity while remaining silent on the mechanism most likely to reduce the resources available for research at all.

***Proportionality***

Section 2.5 established that voluntary norms fail where individual incentives run against collective compliance, and Section 2.7 established that the capacity to absorb compliance costs is unevenly distributed. Proportionality is the constraint that reconciles the two. Governance requirements should scale with the stakes of the particular use, and they should not impose costs that the least-resourced participants cannot meet.

Two design rules follow. The first is that requirements scale with the risk criterion set out above, so that low-risk uses attract a declaration while high-risk uses attract documentation and, where feasible, verification. The second is that the administrative infrastructure required for compliance should be provided centrally rather than assumed to exist. Where elite institutions maintain research integrity offices, smaller universities and institutions in lower-income countries do not, and a mandate that presumes such infrastructure is regressive in effect whatever its intent. Practical implications include machine-readable disclosure templates supplied by publishers rather than composed by authors, funder support for institutional compliance capacity as an eligible cost, and calibration of requirements against cross-country capacity data.

Several institutional pathways are already emerging. The DFG and the ANR are developing funder-level guidelines for AI use in supported research, and the Committee on Publication Ethics has issued guidance on AI and authorship. A reasonable medium-term objective is a coordinated framework at the level of major research councils and scientific associations, analogous to the role the International Committee of Medical Journal Editors plays for clinical trial reporting. Such a framework would set minimum disclosure standards across the four research stages, provide career-stage-differentiated training guidance, and create audit mechanisms for AI-assisted analysis in funded research, while leaving disciplinary norms free to evolve at their own pace.

### 3.3. From diagnosis to instruments

Table 1 summarizes the domains examined in Section 2 and connects each to the governance question, setting the pro-AI case against the governance concern, the key empirical evidence, and the RRAI response. Table 2 then specifies the framework as an implementable schedule, mapping each research stage to the governing principle, the instrument, the actor responsible for implementation, and the indicator by which compliance and effect could be observed. The purpose of Table 2 is to make the framework contestable in detail rather than only in principle.

**Table 1. Generative AI in scientific research: benefits, governance concerns, and the RRAI response**

| Domain | Pro-AI case | Governance concern | Key evidence | RRAI response |
|---|---|---|---|---|
| Research productivity | AI may reverse declining research productivity (Bloom et al., 2020) by reducing the marginal cost of literature synthesis and analysis; expands the hypothesis space through prioritized search (Agrawal et al., 2024) and an effectively larger idea-generating population (Kremer, 1993). | Disaggregation reveals divergence. Publication and citation gains coexist with stagnant novelty and breakthrough output (Park et al., 2023). Homogenization feedback loop: AI outputs become training data, narrowing the knowledge frontier (Shumailov et al., 2024; Doshi and Hauser, 2024). | Publication output up 23.7 to 89.3 percent depending on field and author background; complexity-quality relationship inverted (Kusumegi et al., 2025). Top-decile citation probability up 1.9 percentage points (Chugunova et al., 2026). Throughput up 40 percent, quality ratings up 18 percent (Noy and Zhang, 2023). Topics studied down 4.63 percent, engagement down 22 percent (Hao et al., 2026). | Narrative: assessment frameworks should weight indicators AI does not inflate (replication, data and code availability, external use), so that inflated volume is not read as accelerated progress. Disclosure at analysis stage supplies the data needed to measure the divergence rather than infer it. |
| Researcher selection | If output quality improves, the origin of the insight is secondary. Science exists to produce knowledge, not to test unaided human performance. | Adverse selection under information asymmetry (Akerlof, 1970). If evaluators cannot separate human from machine contribution, past output loses predictive value for future capability, and selection may reward skilled AI users over skilled researchers. | Only 1.5 percent of medical manuscripts disclose LLM use, and revealed use lowers perceived quality by 0.32 points (BaHammam, 2025). Prevalence estimates vary widely by method and corpus (Kobak et al., 2025; Liang et al., 2024). A fully AI-generated manuscript passed workshop review (Lu et al., 2026). Jagged frontier of AI capability (Dell'Acqua et al., 2023). | Disclosure: mandatory, standardized, stage-level and intensity-level, tiered by the probability that an error passes undetected multiplied by the consequence of propagation. Disclosure does not restrict use; it makes the informational environment legible. |
| Workload and funding | AI frees researchers from routine tasks and redirects effort toward conceptual work. General-purpose technologies involve transition costs, and benefits outweighed risks once institutions adapted (Mokyr, 2005). | Net workload rises through competitive escalation and new AI-induced tasks (Grafstrom, 2025). Fiscal corollary: the attribution problem makes research budgets vulnerable, and AI in proposal writing degrades the proposal as a signal. | Verification and correction costs offset production gains in workplace settings (Niederhoffer et al., 2025). AI contribution to total factor productivity of 0.66 percent over a decade (Acemoglu, 2025), consistent with displacement-reinstatement models (Acemoglu and Restrepo, 2019). | Narrative: a shared and accurate formulation that AI redistributes tasks and expands verification work rather than reducing total labor, plus funder statements of what will not be inferred from rising publication volume. |
| Doctoral training | Calculator analogy: previous productivity tools generated skill-loss anxiety and each time the relevant skills adapted. Logical | Counter-analogy: calculators automated procedural tasks, while AI operates on the substance of doctoral work. Mode of engagement matters, | Average learning effect of LLM access is null, but solution-seeking use lowers understanding while explanation-seeking use raises it | Differentiation: career-stage delegation schedules, pseudo-code before functional code, annotated interaction logs, and verification seminars. Guidance rather than |

| Domain | Pro-AI case | Governance concern | Key evidence | RRAI response |
|---|---|---|---|---|
| | thinking and causal inference remain human tasks. | and copy-and-submit is individually rational under publication pressure (Van Quaquebeke et al., 2025). | (Lehmann et al., 2025). Weaker connectivity, recall, and ownership under LLM use in a small neuroimaging study (Kosmyna et al., 2025). Tenure-pipeline simulation shows junior faculty disproportionately harmed (Jiang, 2025). | prohibition, targeting how AI is used and not whether it is used. |
| Self-regulation track record | The governance response was swift: every major publisher issued disclosure requirements within months in 2023. Professional norms are flexible and field-sensitive in a way centralized rules are not. | Publisher policies remain inconsistent (Resnik and Hosseini, 2025), and disclosure rates remain near 1.5 percent because incentives punish disclosure. Pre-registration is the closer analogy than data sharing: voluntary norms fail where individual cost exceeds collective benefit. | 85 percent of researchers want regulatory guidance and 58.7 percent want it from supranational bodies (Chugunova et al., 2026). Voluntary AI guidelines proliferate with almost no binding force (Jobin et al., 2019). FAIR principles irregularly complied with (Wilkinson et al., 2016); pre-registration a minority practice (Nosek et al., 2015). | Institutional governance as a complement rather than a replacement, anchored on binding or bindable instruments above the publisher level (EU AI Act, UNESCO, OECD, national funders) and constrained by proportionality. |
| Attribution and intellectual property | AI as a tool of authorship and invention continues long-standing collaboration patterns, and publisher rules already clarify that only humans can be authors (COPE, 2023). | AI mediation breaks the chain of attribution. Patent law cannot accommodate AI inventorship. The living-research-object scenario removes the stable artifact on which attribution depends, and the private return to original ideas falls (Romer, 1990). | DABUS rulings consistently exclude AI as inventor (Thaler v. Vidal, 2022; Abbott, 2020). Citation indexing presupposes a traceable chain (Garfield, 1979). Sourcing and detection tools emerging (Kobak et al., 2025; Liang et al., 2024). | Disclosure extended to attribution: AI-aware citation standards, mandatory documentation of AI-mediated literature work, and new contribution categories distinguishing original framing from AI-assisted execution. |
| Equity and distribution | AI is an equalizer. It lowers barriers for non-native speakers and resource-constrained researchers, and the counterfactual of no AI means greater concentration, not less (Chugunova et al., 2026). | Adoption is uneven along geography, gender, discipline, and topic, AI infrastructure is consolidated in high-income countries, and the resulting dependencies shape what is knowable and for whom (Edwards, 2010; Jasanoff, 2004; Birhane et al., 2024). | Modest visibility gains for non-Western authors despite higher adoption (Khan et al., 2025). 71 percent of the gender gap explained by familiarity (Chugunova et al., 2026; Carvajal et al., 2024). Median adopted model 26 times smaller than median model built; closed-model adoption growing at 336 percent annually (Trisovic et al., 2025). | Proportionality: requirements scale with stakes, compliance infrastructure supplied centrally rather than assumed, machine-readable templates from publishers, compliance capacity as an eligible funder cost, and calibration against cross-country capacity data. |

| Domain | Pro-AI case | Governance concern | Key evidence | RRAI response |
|---|---|---|---|---|
| Normative fault line | Output-maximizing view: academic institutions exist to produce knowledge, so any tool raising output quality is beneficial and transition costs are offset by long-run gains. | Capacity-development view: institutions also train researchers, certify expertise, and allocate merit. AI substitution introduces costs invisible in productivity metrics, and the homogenization loop undermines the long-run frontier itself. | Illusions of understanding (Messeri and Crockett, 2024). AI may displace human accountability as the voice of science (Silva Luna et al., 2025). Capacity effects observable in learning and verification outcomes (Lehmann et al., 2025). | RRAI adopts the capacity-development view explicitly rather than implicitly, and preserves developmental and certification functions where output-maximizing logic would favor unrestricted adoption. |

*Note. Sections referred to in the table are those of this article. Full citations appear in the reference list.*

**Table 2. RRAI implementation schedule: research stage, principle, instrument, actor, and observable indicator**

| Research stage | Governing principle | Instrument | Implementing actor | Observable indicator |
|---|---|---|---|---|
| Funding | Disclosure | Declaration of AI use within the proposal, disaggregated by component (conceptual framing, methodological design, text preparation), with acceptable-use policy stated ex ante by the funder. | Research councils and national funding agencies (DFG, ANR, ERC) | Share of proposals carrying a declaration; variance in declared use across panels and disciplines. |
| Funding | Proportionality | Compliance capacity funded as an eligible cost; disclosure templates supplied by the funder rather than composed by applicants. | Research councils; supranational bodies for calibration | Take-up of compliance support by institutional income group; ratio of compliance cost to grant value. |
| Research tasks | Disclosure | Stage-level and intensity-level log distinguishing literature review, data analysis, code generation, and drafting, tiered by undetected-error risk. Analysis and code attract documentation; language editing attracts declaration only. | Institutions, research groups, principal investigators | Completeness of logs at project close; audit pass rate for AI-assisted analysis in funded projects. |
| Research tasks | Differentiation | Career-stage delegation schedule; pseudo-code required before functional code; annotated AI interaction logs submitted with written work; methods of AI use section in doctoral theses; verification seminars in curricula. | Doctoral programs and departments | Verification-task performance of cohorts trained under guidance against cohorts trained without it, holding AI access constant. |
| Publication and peer review | Disclosure | Methods of AI use statement for authors, structured on the same tiers. Tiered Reviewer AI Use Statement replacing blanket prohibition, distinguishing language checking, literature retrieval, drafting of comments, and assessment of technical validity. | Journals, publishers, COPE, ICMJE-analogue coordinating body | Disclosure rate by tier; review-outcome differential between disclosing and non-disclosing manuscripts, which tests the normative inversion directly. |

| Research stage | Governing principle | Instrument | Implementing actor | Observable indicator |
|---|---|---|---|---|
| Publication and peer review | Proportionality | Machine-readable disclosure templates supplied by the publisher; no fee-bearing compliance step; no requirement presupposing an institutional research integrity office. | Publishers and editorial offices | Author time to complete disclosure; abandonment rate at the disclosure step by institution type. |
| Use and uptake | Narrative | Agreed formulation on labor effects adopted by professional associations; indicator reform weighting replication, data and code availability, and external use; funder statements specifying what will not be inferred from rising publication volume. | Professional associations, funders, research assessment bodies | Presence of published non-inference statements; weight assigned to non-inflatable indicators in national assessment frameworks. |
| Use and uptake | Differentiation | Written guidance for hiring, tenure, and promotion committees on how to read an AI use statement without penalizing candor. | Universities and evaluation bodies | Share of committees operating with written guidance; direction of the disclosure penalty in appointment outcomes. |
| All stages | Proportionality | Calibration of requirements against cross-country capacity data, with graduated implementation timelines by capacity band. | Supranational bodies (UNESCO, OECD) and monitoring instruments such as the Global AI Index | Ratio of compliance cost to research budget by country income group; dispersion of disclosure rates across capacity bands. |

*Note. Indicators are stated so that each instrument can be evaluated against evidence rather than asserted. Section 3.4 specifies the propositions these indicators would test.*

### 3.4. What would validate or falsify RRAI

Section 2.2 observed that estimates of undisclosed AI use vary widely across detection methods and corpora. That dispersion is also why the framework proposed here cannot yet be evaluated: the research system holds no instrument for measuring its own AI mediation. Tiered disclosure is such an instrument, and that is the first argument for it. A schedule organized by research stage produces panel data on AI use as a by-product of compliance, with no separate collection effort. Most governance frameworks can only be judged long after adoption. This one generates the evidence needed to judge it.

Two of the four principles yield propositions testable in the near term. The first concerns disclosure and the asymmetry it addresses. If the tiering criterion holds, then under a voluntary regime disclosure should be most frequent for low-risk uses and least frequent for high-risk ones, since the reputational cost of admitting AI-generated analysis exceeds the cost of admitting AI-edited prose. Mandating tiers should compress that gradient. No published statistics permit the comparison, because disclosure rates are not disaggregated by research stage, which is why measurement has to precede mandate. A sharper version of the same test concerns the normative inversion. If disclosure can be made to signal rigor, manuscripts carrying a detailed AI use statement should stop attracting a review penalty once journals require and display such statements. Journals already hold the relevant data in their editorial records, and adoption of the Table 2 schedule will be staggered across titles, which supplies a comparison group without anyone running an experiment. It is the cheapest test in the set and the most informative.

The second proposition concerns differentiation and capacity depletion. If mode of engagement is the operative variable, doctoral cohorts trained under delegation guidance should outperform cohorts trained without it on verification tasks, holding AI access constant. Lehmann et al. (2025) implemented this design within a single course. Extending it to a program is a question of scale, not of method.

Narrative and proportionality are harder to test. Both act on institutional behavior instead of individual conduct, and their outcomes, funder inference and the incidence of compliance cost, have no clean counterfactual. What can be observed is uptake: whether funders publish statements of non-inference, and whether the ratio of compliance cost to research budget diverges across capacity bands. We report these as the weaker tests they are.

Two findings would require redesign, not qualification. If mandated tiered disclosure produced no more compliance than voluntary disclosure, the binding constraint would be verification and not requirement, and RRAI would need an enforcement component it does not have. If detailed disclosure continued to attract a penalty after journals displayed it, disclosure could not be made to function as a signal of rigor, and the informational level of the framework would have to be rebuilt around audit instead of transparency.

The identification problems are plain. Journals adopting tiered disclosure early will not be a random sample, doctoral programs adopting delegation guidance will differ from those that do not, and researchers who disclose will differ from those who do not along dimensions the disclosure does not record. None of this makes the propositions untestable. It does mean the first move belongs to journals, which already hold the data and face the lowest cost of adoption.

## 4. Concluding remarks

This paper has moved beyond the individual effects of generative AI in scientific research to ask whether the existing self-regulatory apparatus of the academic system can manage the collective consequences of widespread adoption. The disaggregated productivity picture shows individual gains coexisting with ambiguous trends in novelty, disruption, and breakthrough output. The adverse selection problem, the workload paradox, the bootstrapping challenge in doctoral training, the structural inequality in access to AI infrastructure, the erosion of attribution, and the homogenization feedback loop together indicate that the transition to AI-assisted research carries systemic risks that individual researchers can neither observe nor manage on their own.

We have argued that these risks are not a single problem. They arise through three distinct mechanisms, and the governance instrument required differs accordingly. Disclosure addresses information asymmetry, differentiation addresses capacity depletion, narrative addresses the fiscal externality of misrepresented labor effects, and proportionality constrains all three so that the burden of compliance does not fall where capacity is weakest. The choice facing the research community is not between unrestricted adoption and top-down regulation. It is between designing these instruments deliberately and inheriting them from whichever actor moves first.

RRAI is offered as one architecture rather than the only one, and we expect it to be refined, contested, and improved through scholarly and institutional engagement. That engagement should run across all four stages of the research process, from funding to use, and should include the communities most exposed to the distributional consequences of AI adoption: researchers in lower-income countries, early-career researchers, and the wider public affected by what science does and does not investigate.

### Declaration of AI use

In keeping with the disclosure architecture proposed in Section 3, we report our own AI use according to the tiered schedule set out in Table 2.

Language editing (low tier): Anthropic's Claude was used for copy-editing and for improving clarity of expression throughout. Literature synthesis (intermediate tier): Claude was used to assist in identifying and summarizing relevant literature. Every reference cited in this article was independently verified by the authors against the published source. Drafting (intermediate tier): Claude was used to produce and revise draft text, which the authors then edited substantively. Data analysis and code generation (high tier): not applicable. This article contains no empirical analysis. The authors reviewed and edited all content and take full responsibility for the article, including the accuracy of all citations and of all statistics attributed to cited sources.

### ORCID

Fulvio Castellacci: 0000-0001-7707-7397
Tommaso Ciarli: 0000-0003-0362-8496
Yuan Gao: 0000-0003-3505-9304
Marianna Marino: 0000-0003-2712-8664
Giacomo Marzi: 0000-0002-8769-2462
Massimo Riccaboni: 0000-0003-4979-8933
Maria Savona: 0000-0003-4529-4869
Simone Vannuccini: 0000-0001-7083-6178

### Declaration of competing interest

The authors declare that they have no known competing financial interests or personal relationships that could have appeared to influence the work reported in this paper.

### Funding

This research did not receive any specific grant from funding agencies in the public, commercial, or not-for-profit sectors.

### Acknowledgments

The authors thank the participants at the AI for Science and Innovation Workshop held at the IMT School for Advanced Studies Lucca on April 8 and 9, 2026. The discussion sparked at the final roundtable are represented in this commentary.